Short Paper

# Effectiveness of an Online Course in Programming in a State University in the Philippines

Aaron Paul M. Dela Rosa
College of Information and Communications Technology, Bulacan State University, Philippines
ORCID: 0000-0003-1662-083X
aaronpaul.delarosa@bulsu.edu.ph

**Abstract**

*Purpose* – Online courses, as a pedagogical approach to teaching, boomed during this Coronavirus Disease 2019 (COVID-19) pandemic era. Universities shifted from traditional face-to-face classes to online distance learning due to the cause of the pandemic. This study aimed to determine how effective an online course is in learning a programming course.

*Method* – The study utilized mixed-method research applied through a validated survey questionnaire consisting of closed- and open-ended questions. Python programming was the course selected to undergo the study and underwent an evaluation to determine the students' responses. Student respondents are from Bulacan State University, a state university in the Philippines, under the Bachelor of Science in Information Technology program.

*Results* – Based on their responses, the students found that the online Python programming was Very Effective, with an overall mean of $\bar{x}=4.49$. This result shows that students found the online course effective, provided the proper course design and content, allowed them to spend enough time finishing tasks, and provided communication and interaction with their instructor and fellow students. Additionally, students gave overwhelmingly positive responses when asked what their instructors had done well on the course delivery and provided insightful and constructive comments for further enhancement and delivery of the course.

*Conclusion* – This study found that most students strongly agreed and believed in the effectiveness of delivering the Python Programming course asynchronously. With such positive results from the student's perspective and evaluation, the course can be enhanced to continue providing quality education at Bulacan State University.

*Recommendations* – Revisit the course syllabus to identify the contents that should be well interconnected. Include more examples, exercises, quizzes, and activities throughout the course to help students assess their learning. Provide more means of communication and interaction with students even if the course is delivered asynchronously.

*Research implications* – Considering its results, this study shows that programming courses delivered in higher education institutions could be delivered successfully through online learning, with asynchronous learning modality, and that faculty members can still intervene with students' learning to address questions and clarifications.

---

## INTRODUCTION

Coronavirus Disease 2019 (COVID-19) pandemic started in December 2019 and began its spread throughout the globe. With its effect, businesses, offices, and even schools and universities were forced to shut down and find alternative ways to reach their students (Khalil et al., 2020). Schools and universities encountered challenges transitioning from traditional face-to-face classes to online distance learning (Mpungose, 2020; Turnbull et al., 2021).

The Republic of the Philippines was not exempted from the spread of the pandemic. The government, as a solution addressing the global concern on education, found its way to deliver learning amidst the pandemic, with different learning modalities being imposed (CMO No. 4, s. 2020, 2020). One of the state universities in the Philippines, Bulacan State University, developed guidelines in accordance with the government solution. Bulacan State University (BulSU) is a progressive knowledge-generating institution located in the City of Malolos, Bulacan, Philippines. In its efforts to provide education to its students, BulSU developed guidelines for implementing flexible learning modalities (Bulacan State University [BulSU], 2020). BulSU implemented three (3) learning modalities during this pandemic: (1) Synchronous Learning (SL), (2) asynchronous online learning (AOL), and (3) Remote-print Learning (RPL). These learning modalities were implemented to accommodate the diverse needs of its students.

One of the colleges at BulSU, the College of Information and Communications Technology (CICT), is implementing AOL as its learning modality throughout all courses offered within the college. Through AOL, learning materials and classwork are given to the students beforehand to have ample time to read and study the materials. Afterward, the faculty members will conduct a synchronous session for consultation with the students.

As Eusoff et al. (2022) mentioned in their study, delivering programming courses online is a challenge for educators, as they have fewer interactions with the students, not knowing if the students fully understand the concept, the context, and the logic throughout a programming course since programming is a complex course to learn. True enough, CICT experienced the same concern while teaching programming. With such concern, this study is focused on determining the effectiveness of an online course in teaching and learning Python programming offered by BulSU-CICT.

This study aims to answer the following research questions:

1. How may the effectiveness of the online course in Python programming be determined in terms of navigation and time on task organization, course design, communication and interaction, and course content?
2. How may the students describe their perception of what the instructor has done well and possible room for improvement in the course?

## LITERATURE REVIEW

### *Online Learning*

In terms of how the students perceived online learning, Stone (2017), Coman et al. (2020), Khalil et al. (2020), Tathahira (2020), and Laili and Nashir (2021) highlighted in their studies on how students gain an advantage on online learning and how to improve their participation in the course. Though delivered online, online courses shall not be far from the traditional learning approach, where students shall be given enough resources and ample time to complete tasks. Students' perception of an online course is essential since they are the primary users of it.

The researcher focused on the students' perspectives on evaluating an online course in programming. The evaluation presents the online course's navigation and time-on-task organization, course design, communication and interaction, and course content. Students evaluated the course based on these criteria to determine how they perceived the delivery of the online course in programming. Additionally, students were given the freedom to mention how well the instructor conducted the course and how it could be improved for further use of the online course.

### *Online Learning During the COVID-19 Pandemic*

As higher education institutions shift from traditional face-to-face classes to online learning due to the effects of the COVID-19 pandemic, these institutions found a way to continue to provide quality education amidst the challenges facing them. Several studies have highlighted how universities handled this situation by implementing online learning. Abdulkareem and Eidan (2020), Simamora et al. (2020), Coman et al. (2020), Laili and Nashir (2021), and Ali (2020) studied the effects of the pandemic in higher education institutions, highlighting the challenges encountered. Considering the situation, these authors have emphasized how universities could adapt to continue to provide education through online learning. These studies mentioned the advantages of online learning amidst the pandemic. Moreover, they have found that within the situation, online learning became the only resort to continue delivering education to its people.

Online learning seems to have found its way to being the solution during this pandemic. The researcher maximized the use of online learning and evaluated a programming course delivered asynchronously. This presents the perspective of the students on taking a programming course online.

### *Learning Python Programming in an Online Learning Environment*

Online courses can be supplementary when in higher education, specifically in the field of information technology, where several available resources can be found on the internet as an additional tool in learning a programming language. These studies have highlighted how Python programming can be delivered online and how it affects the knowledge and skills in programming. Bustos and Álvarez-González (2021), Moumoutzis et al. (2018), and Sharp (2019) have studied how Python programming can be delivered using existing tools online. These studies found that it fulfilled several skills of the students while taking the course online.

This study focused on evaluating the effectiveness of the Python programming course delivered asynchronously, where materials and classwork are given to students in advance, and an intervention by the instructor is conducted through synchronous sessions. The researcher focused on Python programming as it is one of the emerging programming languages and popular in industries from the Philippines and even around the globe. Students were asked to evaluate the course at the end of the semester, after taking all modules within the online course.

## METHODOLOGY

### *Research Design*

The study used mixed methods research, an approach to analyze and interpreting both quantitative and qualitative data (Natividad-Franco, 2022; Bautista, 2023; Natividad-Franco et al., 2023). As Halcomb and Hickman (2015) stated, mixed-method research enables the collection of both quantitative and qualitative data at the same time, reducing the effort in data collection.

According to Basias and Pollalis (2018), quantitative research "involves systematic and empirical investigation... through statistics and mathematics and the processing of numerical data." Respondents were given a close-ended questionnaire that could be rated from one to five and then further analyzed to present the numerical data. On the other hand, qualitative research examines experiences and behaviors without statistics and numerical data. The study covered two open questions to its respondents about how their instructor did well in the course and the possible room for improvement.

### *Research Instrument*

The primary source is the evaluation of online courses/teaching in the Department of Clinical Sciences of Colorado State University, which was prepared by Stewart and Kogan (2015). Then this evaluation was made to be a survey questionnaire that was validated by the BulSU Teaching and learning technologies experts of BulSU. The survey questionnaire consisted of both closed and open-ended questions. Closed-ended questions were analyzed using descriptive statistics, presenting mean and verbal interpretations. Open-ended questions were analyzed using qualitative content analysis to generate patterns from the textual responses (Lindgren et al., 2020; De Brún et al., 2022).

### *Data Gathering Procedure*

The data collection began after the students completed the course. The survey questionnaire developed from the evaluation of online courses/teaching was included at the end of the course for the student to evaluate. The questionnaire was attached at the end of the course in a separate module using the Canvas Learning Management System (LMS), which was used as the LMS for teaching and learning Python programming.

### *Population and Sample*

The students involved in this study are the Bachelor of Science in Information Technology (BSIT) 4th-year students who took the Python Programming course. A total of 220 students answered the survey who were enrolled in the course during the first semester of the academic year 2021-2022.

### *Statistical Treatment*

The interpretation was based on a five-point Likert scale in terms of effectiveness. Table 1 presents the range, scale, and descriptive interpretation.

Table 1. Five-Point Likert Scale

| Scale | Range | Descriptive Rating |
|---|---|---|
| 5 | 4.50 – 5.00 | Extremely Effective |
| 4 | 3.50 – 4.49 | Very Effective |
| 3 | 2.50 – 3.49 | Effective |
| 2 | 1.50 – 2.49 | Somewhat Effective |
| 1 | 1.00 – 1.49 | Not Effective |

## RESULTS AND DISCUSSION

### *Effectiveness of the Online Course in Python Programming*

The evaluation tool used in this study was developed from the evaluation of the online course/teaching from Colorado State University (Stewart & Kogan, 2015). The instrument evaluates an online course regarding Navigation and Time on Task Organization, Course Design, Communication and Interaction, and Course Content. Tables 2 to 5 present the students' evaluation of the different criteria. Table 6 summarizes the overall rating of the students' evaluations of the online course.

Table 2. Responses in terms of Navigation and Time on Task Organization

| Item | Mean | Descriptive Interpretation |
|---|---|---|
| Navigational instructions make the organization of the course easy to follow. | 4.43 | Very Effective |
| Provides orientation to the course and its structure. | 4.55 | Extremely Effective |
| Clearly organizes and explains online assignments and related due dates. | 4.49 | Very Effective |
| Uses modules to organize course content. | 4.55 | Extremely Effective |
| Clearly presents expectations and grading policies. | 4.53 | Extremely Effective |
| **Total** | **4.51** | **Extremely Effective** |

Navigating through the course and time allotted on task organization has been found to be "Extremely Effective" with a total mean of $\bar{x}$=4.51. This shows that students using this online course quickly found their way through it and are less likely to be misled while navigating from one content to another.

Table 3. Respondents' Ratings in terms of Course Design

| Item | Mean | Descriptive Interpretation |
|---|---|---|
| Online course design clearly articulates course policies and procedures. | 4.43 | Very Effective |
| The learning modules clearly state the learning goals. | 4.46 | Very Effective |
| The course uses a variety of online tools to facilitate student comprehension and engagement. | 4.46 | Very Effective |
| Online course content addresses different learning styles. | 4.40 | Very Effective |
| Online course design describes available technical support. | 4.40 | Very Effective |
| Communicates a sense of excitement and excitement. | 4.30 | Very Effective |
| **Total** | **4.41** | **Very Effective** |

The online course's design was referenced from the course syllabus of the select course. By triangulating the structure of the syllabus with the online course, the course design was considered 'very effective' with a total mean of $\bar{x}$=4.41. Students could determine which module they should be in each week based on how the online course was designed. This has allowed students to be guided appropriately even if different instructors teach them.

Table 4. Responses in terms of Communication and interaction

| Item | Mean | Descriptive Interpretation |
|---|---|---|
| The instructor responds to emails within 72 hours or less. | 4.35 | Very Effective |
| Encourages mutual respect among students. | 4.49 | Very Effective |
| Encourages students to interact with each other and with the instructor. | 4.48 | Very Effective |
| Treats class members equitably and respectfully. | 4.56 | Extremely Effective |
| Respond constructively to student questions, opinions, and other input. | 4.55 | Extremely Effective |
| Recognize and respond when students do not understand. | 4.50 | Very Effective |
| Creates a sense of community in the online course. | 4.44 | Very Effective |
| Effectively handles inappropriate discussion postings or other unacceptable online behavior. | 4.53 | Extremely Effective |
| **Total** | **4.49** | **Very Effective** |

Students are deemed more interactive in the online course with proper guidance and communication with their instructor. Thus, in terms of communication and instruction, it has been found to be "Very Effective," with a total mean of $\bar{x}$=4.49. Although the content was provided well to the students, guidance of the instructor tends to the students to be more interactive in class, through the online course, or through synchronous sessions. Students found it effective when the instructor treated each of them equally, respected them, and paid attention.

Table 5. Respondents' Ratings in terms of Course Content

| Item | Mean | Descriptive Interpretation |
|---|---|---|
| Demonstrates appropriate depth of knowledge of course subject. | 4.57 | Extremely Effective |
| The content is appropriate for the course level. | 4.59 | Extremely Effective |
| Explains difficult terms, concepts, or problems in more than one way. | 4.51 | Extremely Effective |
| Relates assignments to course content. | 4.56 | Extremely Effective |
| Includes examples relevant to student experiences and course content. | 4.54 | Extremely Effective |
| Provides opportunities for students to engage in active learning. | 4.51 | Extremely Effective |
| **Total** | **4.55** | **Extremely Effective** |

Students tend to learn better with multiple examples given to them to understand a lesson better, specifically in programming, which is a course that requires students to apply what they learned. Thus, the online course's content was deemed "Extremely Effective" with a mean of $\bar{x}$=4.55. This shows that the course content provides all the necessary topics for the Python programming course and provides assessments that would hone the student's knowledge of a particular lesson.

Table 6. Overall Summary of the Respondents' Ratings in Evaluating the Online Course

| Item | Mean | Interpretation |
|---|---|---|
| Navigation and Time on Task Organization | 4.51 | Extremely Effective |
| Course Design | 4.41 | Very Effective |
| Communication and Interaction | 4.49 | Very Effective |
| Course Content | 4.55 | Extremely Effective |
| **Overall Mean** | **4.49** | **Very Effective** |

Focusing on its strength, students found the course content was effective, allowing them to learn better through the materials, classwork, and assessments provided within the course. Additionally, the time allotted to complete the tasks is sufficient for students to answer. In summary, the student's responses to the online course have been found to be "Very Effective," with an overall mean of $\bar{x}$=4.49.

## *Students' Perceptions of Course Delivery by the Instructors and Room for Improvements*

The survey questionnaire consisted of two (2) open-ended questions for students to evaluate their instructors' course delivery during the semester and what their instructors could do more to improve the course. Among the 220 student respondents, 218 answered open questions. A word cloud was generated using Microsoft Power BI. Figure 1 shows the word cloud regarding the first question that tackles how well the instructors' course delivery is. Figure 2 shows the word cloud on the students' perception of what the faculty could do more to improve course delivery.

The Microsoft Excel file of the responses was imported into Power BI to generate the word clouds presented in Figures 1 and 2. Stop words are removed for better analysis of the word cloud, such as 'a', 'an', 'the', 'in', 'at', and other common stop words. Additionally, a manual analysis of the responses was conducted to the responses.

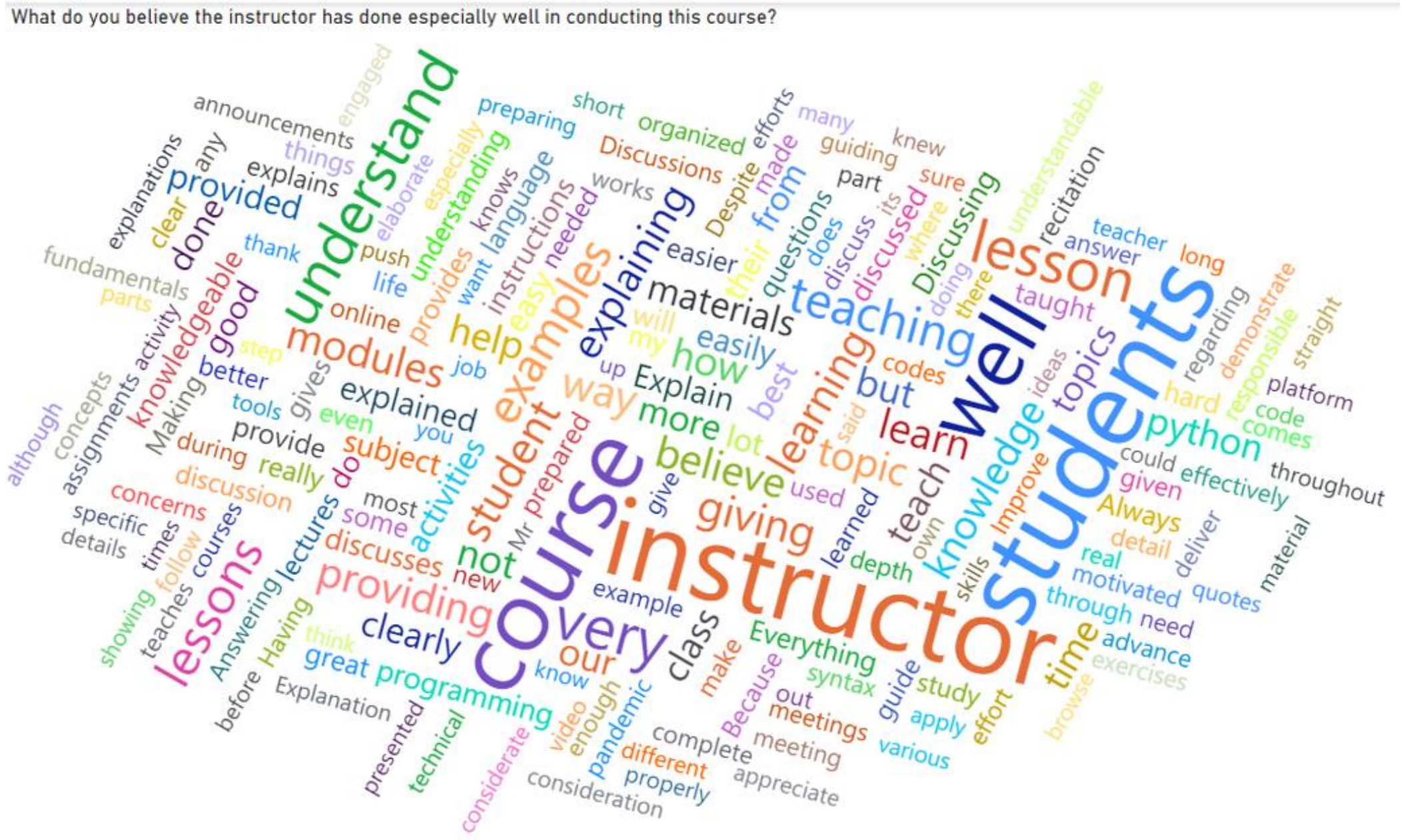

*Figure 1.* Student perception of instructors' course delivery

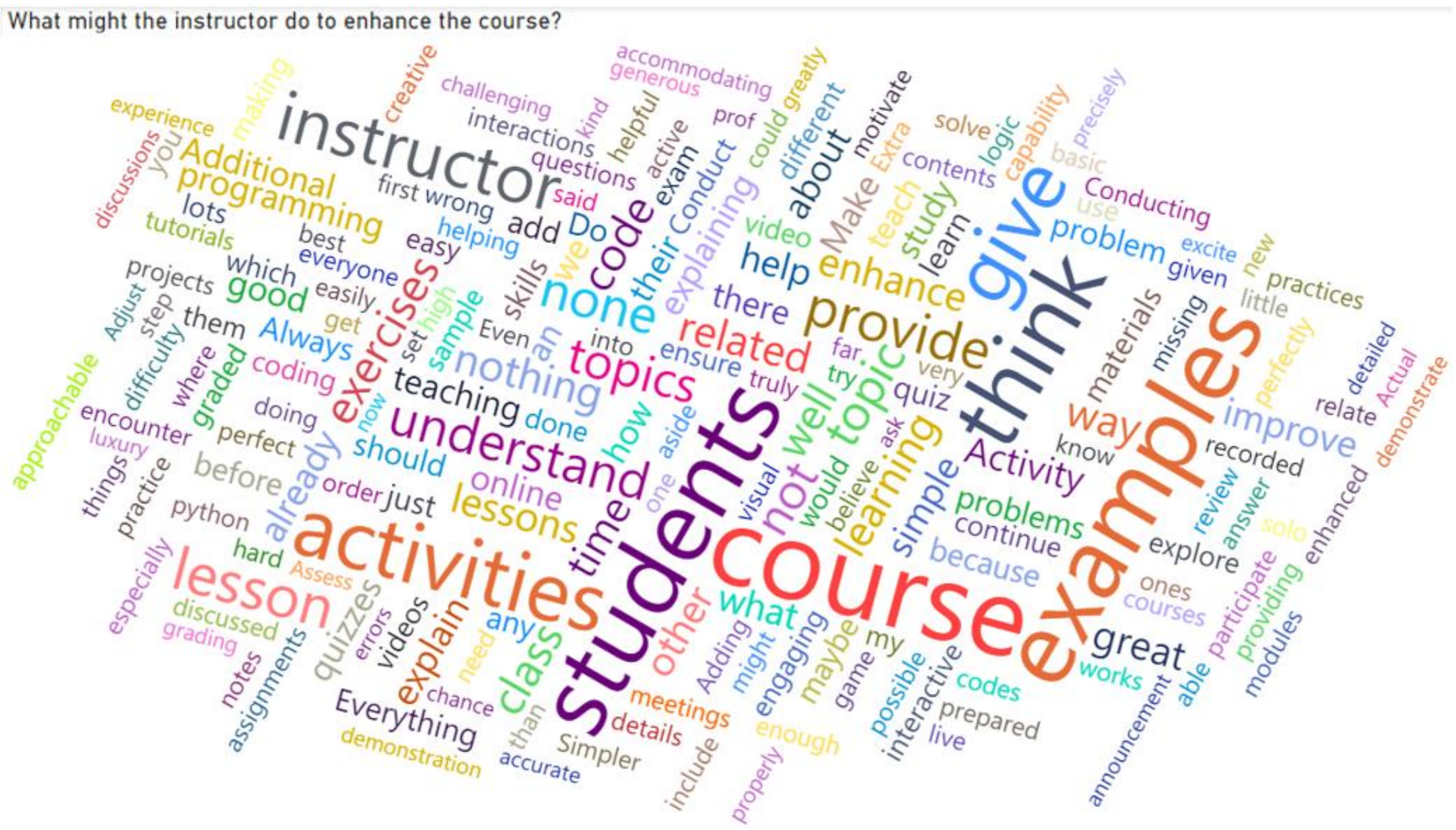

*Figure 2.* Student perception on instructors' enhancement to the course

Regarding the perception of what the instructor did well in conducting the course, the students gave an overwhelmingly positive response to how their instructors delivered the course to them. To the students, the instructors of the course were sufficiently knowledgeable that enabling them to discuss the topics further. Additionally, the students were thankful for the preparation of the materials and how well they organized them. Lastly, the students highlighted that the faculty understands their need to learn the language well by explaining things further and giving more examples.

On the other hand, the students gave comments, suggestions, and recommendations on how the instructors could improve the course. First, students tend to look for more examples during discussions. Others are looking for weekly quizzes and activities for them to be able to test their capabilities and learning to the lesson discussed. Lastly, students are looking for more accommodation since the course was delivered online, and communication is one of the most challenging situations. On the contrary, 30% or 66 students have mentioned that the course is delivered well and did not give other recommendations for enhancements.

## CONCLUSIONS AND RECOMMENDATIONS

After analyzing and interpreting the data, this study found that most of the students strongly agreed and believed in the effectiveness of delivering the Python programming course asynchronously. With such positive results from the perspective and evaluation, the course can be enhanced to continue delivering quality education at Bulacan State University. In addition to this, it was also found that: (1) navigation on the course and the time on task organization was presented and done well, (2) the students were engaged through the course that was referenced from its course syllabus, (3) communication and interaction with their instructor are essential to students taking the online course asynchronously, (4) the course content presented all needed lessons and topics for the students to learn and be engaged with the course, (5) students gave an overwhelmingly positive response to how their instructors conducted and delivered the course, and (6) students gave recommendations to the enhancement of the course.

Considering the conclusions, some recommendations may be included to enhance the course. Revisit the course syllabus to identify contents that should be well-interconnected. Moreover, include more examples, exercises, quizzes, and activities throughout the course for students to assess their learning. Additionally, provide more means of communication and interaction with the students even if the course is delivered asynchronously. Lastly, other courses offered online by the College may have the same delivery and implementation.

## PRACTICAL IMPLICATIONS

Considering the results of the study, it has been found effective that an online course in Python programming enables students to learn the language through asynchronous online learning modality as implemented in Bulacan State University. Faculty member interventions, along with the asynchronous mode, strengthened the students' learnings through thorough and further discussions and clarifications of the module provided within the Python online course. Therefore, other programming courses offered at the university may implement the same approach.

## ACKNOWLEDGEMENT

This research does not receive funding. The author would like to acknowledge the BSIT 4th-year students of BulSU-CICT batch 2022 for participating in this research as respondents and also for recognizing the family and friends of the author.

## DECLARATIONS

### *Conflict of Interest*

The author declares that there is no conflict of interest.

### *Informed Consent*

All participants of the study were informed of the purpose and data to be collected upon answering the survey questionnaire. The identity of the respondents was not obtained during data gathering.

### *Ethics Approval*

The BulSU Research Management Office and the Research Ethics Committee accepted and approved the conduct of the study.

## Author Biography

Aaron Paul M. Dela Rosa is a college instructor at the College of Information and Communications Technology (CICT) of Bulacan State University (BulSU). He presents and publishes research papers at national and international conferences and journals focusing on web application development and related fields. He is a member of multiple national and international organizations, a certified data protection officer, and certified in various programming languages.